# Electron FLASH Delivery at Treatment Room Isocenter for Efficient Reversible Conversion of a Clinical LINAC

Mahbubur Rahman*[1], M. Ramish Ashraf*[1], Rongxiao Zhang[1,2,3], Petr Bruza[1], Chad A. Dexter[3], Lawrence Thompson[3], Xu Cao[1], Benjamin B. Williams[1,2,3], P. Jack Hoopes[1,3,4], Brian W. Pogue[1,3,4], David J. Gladstone[1,2,3]

[1] Thayer School of Engineering, Dartmouth College, Hanover NH 03755, US

[2] Department of Medicine, Geisel School of Medicine, Dartmouth College Hanover NH 03755 USA

[3] Norris Cotton Cancer Center, Dartmouth-Hitchcock Medical Center, Lebanon, NH 03756 USA

[4] Department of Surgery, Geisel School of Medicine, Dartmouth College, Hanover NH 0375 USA

* Authors contributed equally.

**Abstract**

**Purpose:** In this study, procedures were developed to achieve efficient reversible conversion of a clinical linear accelerator (LINAC) and deliver electron FLASH (eFLASH) or conventional beams to the treatment room isocenter.

**Material & Methods:** The LINAC was converted to deliver eFLASH beam within 20 minutes by retracting the x-ray target from the beam's path, positioning the carousel on an empty port, and selecting 10 MV photon beam energy in the treatment console. Dose per pulse and average dose rate were measured in a solid water phantom at different depths with Gafchromic film and OSLD. A pulse controller counted the pulses via scattered radiation signal and gated the delivery for preset pulse count. A fast photomultiplier tube-based Cherenkov detector measured per pulse beam output at 2 ns sampling rate. After conversion back to clinical mode, conventional beam output, flatness, symmetry, field size and energy were measured for all clinically commissioned energies.

**Results:** Dose per pulse of 0.86 +/- 0.01 Gy (310 +/- 7 Gy/s average dose rate) were achieved at isocenter. The dose from simultaneous irradiation of film and OSLD were within 1%. The PMT showed the LINAC required about 5 pulses before the output stabilized and its long-term stability was within 3% for measurements performed at 3 minutes intervals. The dose, flatness, symmetry, and photon energy were unchanged from baseline and within tolerance (1%, 3%, 2%, and 0.1% respectively) after reverting to conventional beams.

**Conclusion:** 10 MeV FLASH beams were achieved at the isocenter of the treatment room. The beam output was reproducible but requires further investigation of the ramp up time in the first 5 pulses, equivalent to <100 cGy. The eFLASH beam can irradiate both small and large subjects in minimally modified clinical settings and dose rates can be further increased by reducing the source to surface distance.

## I. Introduction

There has been a resurgence of interest in delivering radiation treatment at ultra-high dose rates (FLASH, >40 Gy/s) for improved normal tissue sparing while ensuring comparable tumor control to conventional (CONV) dose rates (~0.1 Gy/s). Preclinical studies into high dose rate effects on biological outcomes date back to Hornsey et al,[1,2] when their results showed reduced toxicity in mice at high dose rates. In the last decade, more studies on mice showed FLASH irradiation can reduce late lung fibrosis and acute pneumonitis.[3] Montay-Gruel et al. showed that there are long term neurocognitive benefits of FLASH including maintaining extinction memory and preventing radiation induced depression, anxiety, and neuroinflammation.[4] FLASH treatments were also delivered to other animals including zebrafish, cat and pigs.[5,6] Bourhis et al recently treated the first human patient with the Oriatron eRT6 specifically designed to deliver low energy FLASH electron beams.[7]

Technology and machines capable of delivering FLASH dose rates are available, but in several different forms. X-ray tubes can deliver FLASH dose rates although very superficially (<2 mm depth, 160 KV x-ray tubes).[8] Proton FLASH beams were developed recently to test the FLASH effect on small animals and for preclinical studies.[9–11] Synchrotron microbeam radiation therapy inherently delivers beams via spatial micro fractionation and at extremely high dose rates of (~16 kGy/s).[12] However, for translation into the clinic, the method of delivery requires new ways to understand and prescribe the treatment dose. Absolute dosimetry, treatment planning software (TPS), and quality assurance (QA) of these beams are still in development. Alternatively, conventional clinical linear accelerators (with QA/TPS technology developed) can be modified to deliver FLASH dose rates. Schüler et al and Loo et al modified a Varian Clinac 21EX to deliver FLASH beams and study biological outcome on mice.[13,14] Lempart et al modified and tuned an Elekta LINAC to deliver FLASH beams for future radiobiological experiments.[15] However, FLASH dose rates were achieved by both LINACS inside the gantry head or at less than 53 cm from the target, respectively. This reduced distance severely limits use of FLASH radiation delivery to large animal preclinical studies and treatment of patients.

In this study, a Varian Clinac 2100 C/D (Palo Alto, CA) was modified to deliver FLASH dose rates at treatment room isocenter (100 cm from conventional target). Procedures and guidelines were developed for the conversion of a LINAC to deliver FLASH dose rates to the isocenter or efficiently reversed to deliver conventional beams. The dosimetry techniques characterized the eFLASH beam delivery at submillimeter spatial and single pulse temporal resolution (360 Hz pulses, at 2 ns sampling rate). Dose rates exceeding 300 Gy/s at isocenter were verified with multiple dosimetric modalities.

## II.     Material & Methods

| Step | Description | Accelerator Part Affected |
|---|---|---|
| 1 | In the treatment console, service mode was selected. A **carousel port cover** that would be removed from the **carousel** was identified (e.g. 20 MeV, see figure 1c). The energy of the port OPPOSITE of the identified port cover on the carousel was selected in service mode. | Carousel |
| 2 | The **air drive** was turned off in the **stand.** | Air Valve |
| 3 | **Lead shielding** was removed from the front of the gantry head, so the **port cover** that was to be removed could be visible through the service port. The cover was removed to create an open port. | Carousel |
| 4 | The **carousel** position was manually changed, so electron beams could traverse through the empty port. (Port for beam energy set in service mode for step 1 was visible through the service port) | Carousel |
| 5 | **Lead shielding** was removed from the side of the gantry, so the **target actuator** was visible. Target actuator position was manually adjusted to electron mode (see figure 1a). The **actuator** was restricted from drifting to x-ray modes. | Target |
| 6 | The **energy switch** was verified to be in the position for desired beam energy. Drifting of the position was physically restricted. | Energy Switch |
| 7 | In service mode, **dose, dose-rate, and beam steering servos** were turned off. The desired photon beam energy was selected (e.g. 10 MV) to deliver electron beams. | Energy Switch and Servos Circuits |

**Table 1.** Steps for modifying the LINAC from conventional x-ray photon beams to a FLASH electron beam. Components altered for the upgrade are bolded.

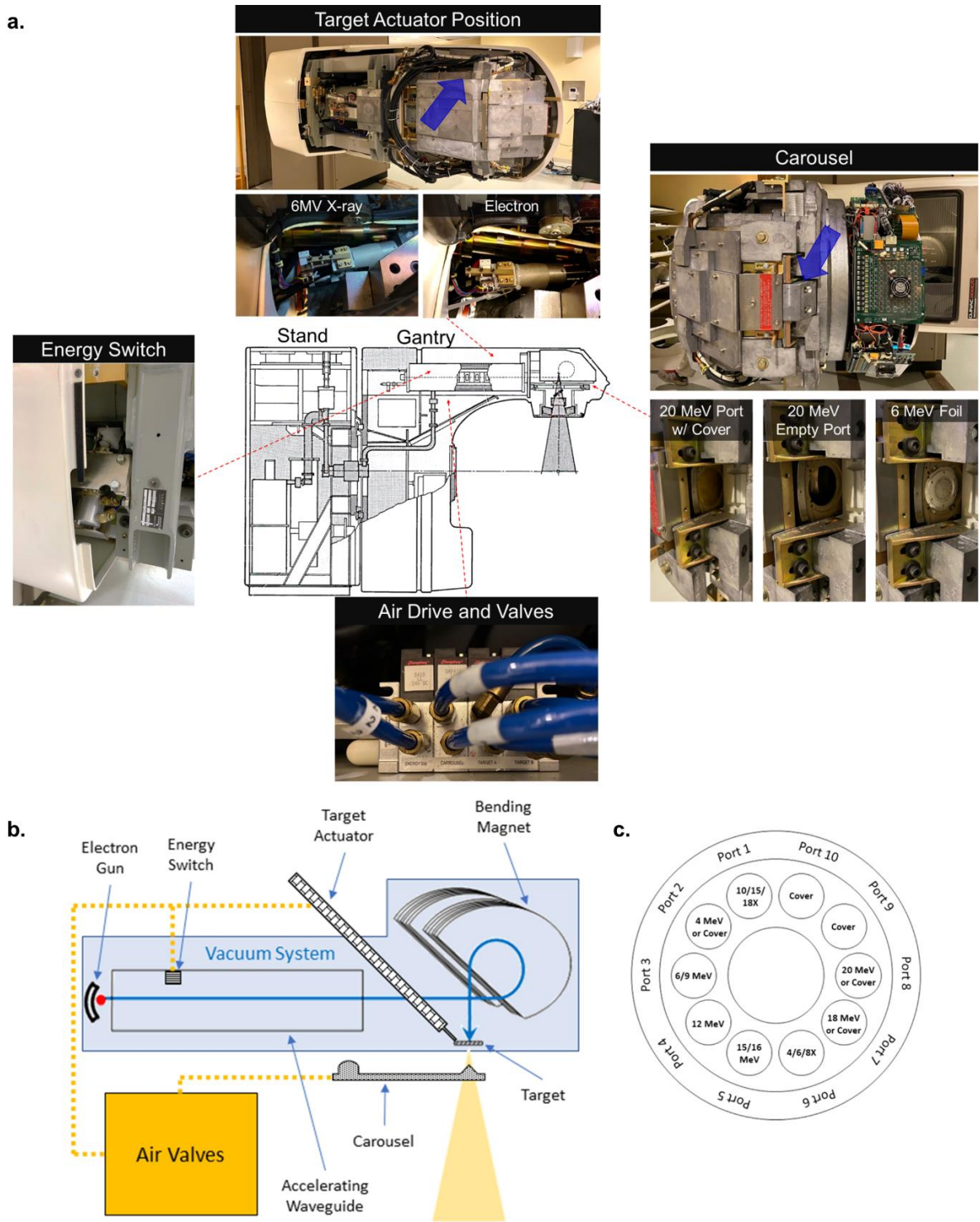

**Figure 1. a.** A high-energy radiotherapy LINAC highlighting the components altered for upgrade to FLASH dose rates. Components include carousel, air valve, target actuator, and the energy switch. The lead shielding that were withdrawn to access the target actuator and the carousel are indicated using the blue arrows. **b.** Schematic diagram of modified components within the LINAC. **c.** Carousel positions.[16,17]

## A. Machine Modifications

The steps to modify the LINAC were included in Table 1. The modifications were done for the LINAC to deliver high fluence (10 MeV) electron beams by choosing a 10 MV photon beam in the treatment console and then removing the target and flattening filter from the beam's path. The modifications (completed within 20 minutes) required manual setting of a few key components of the treatment delivery system: carousel, air valve, target, and the energy switch. The gantry angle was initially set to 90 degrees (horizontal) to access the components. Each component's impact on achieving FLASH dose rates, dose, dose-rate, and beam steering servos, interlocks, and conversion back to clinical conditions are described below.

### 1. Carousel

The carousel of the treatment head contained flattening filters (FF) and scattering foils (SF) for each commissioned clinical energy (Fig 1c). Flattening filters and scattering foils are utilized in conventional radiation therapy for homogeneous dose distribution in x-ray mode and electron mode, respectively.[18] However, flattening filters and scattering foils reduce the dose rate (by 2-4 times, and ~30 respectively at treatment room isocenter).[19,20] Typically, the carousel has unused positions with port covers inserted. One such position was chosen, and the port cover was removed resulting in an open port. The access to the port cover was realized by removing a small lead shield block from the front of the gantry (Fig 1). To access the carousel port cover for removal, first we selected a beam energy corresponding to the opposite carousel position (see Fig 1c). Specifically, in Step 1 (Table 1) of the modification process, a 6 MeV beam was chosen in service mode, such that the 20 MeV port cover was accessible through the shield opening. In Step 2, pneumatic drive was overridden to prevent inadvertent rotation of the carousel (see next paragraph), followed by manual port cover removal (Step 3). After removal of the 20 MeV cover, the carousel was manually rotated to the opposite position (Step 4) and shown in Figure 1a.

### 2. Air Drive

The entire electron beam transport system from the electron gun, through the accelerating waveguide, bending magnet, and target, to the exit window is under vacuum to prevent scatter of the electron beams (Fig 1b).[16,21] In conventional LINAC operation, mechanical adjustments of the components are required to select the beam energy within the vacuum system, including positions of the energy switch, target, and carousel. Mechanical adjustments are pneumatically actuated and controlled by air valves (Fig 1a). By turning off the air drive in step 2, the LINAC can be manually adjusted to deliver high intensity electron beams (via photon mode). Furthermore, it prevents the system from rotating the carousel and changing target positions to produce x-ray beams. The energy switch and the target are vacuum sealed and pressure differences can cause both components to drift and prevent delivery of the high intensity electron beams. Therefore, when the appropriate energy switch and target position have been chosen (discussed in iii and iv), drifts were avoided by mechanically preventing the energy switch and the target actuator to change position.

3. **Energy Switch**

The energy switch changes the ratio of the microwave power into the first portion (electron gun end) and second portion (output end) of the standing wave accelerator structure by mechanically switching the position of a plunger. In one position there is a strong coupling between the first and second portion of the standing-wave structure for a strong electric field and delivery of high energy x-ray beams (e.g. 10 MV, 18 MV etc). In the second position, there is a weak coupling between the two portions for a weak electric field and delivery of low energy x-ray beams (e.g. 4 MV, 6 MV).[16] In this study, the energy switch was in the first position for delivery of a high dose-rate 10 MeV beam with beam current corresponding to conventional 10 MV x-ray beam.

4. **Target**

The target actuator and positioning mechanism (shown in Fig 1a) had three modes: low energy x-ray mode (e.g. 6 MV), high energy x-ray mode (e.g. 10 MV) and electron mode. The target in conventional x-ray modes converts the high intensity electron beams into x-ray beams via Bremsstrahlung at approximately 1-10% efficiency.[22] When the target was removed from the beam path the high electron beam fluence intended for photon production was transmitted through the accelerator structure and delivered a high fluence FLASH beam for experimental use. The target mechanism was manually pulled to electron mode for delivery of FLASH electron beams, and temporarily fixed to prevent it from being pushed back due to atmospheric pressure. The gantry angle was then set to 0 degrees with the beam directed towards the ground.

**5.     Dose, Dose-Rate, and Steering Servo**

In conventional clinical settings, two ionization chambers within the gantry head monitor the beam's dose, dose rate (both dose per pulse and average dose rate), symmetry, positioning. These parameters are controlled by the LINAC through feedback servo circuits to ensure the beam shape and intensity is within a specific tolerance of normal conditions. However, after modifying the LINAC the servos would expect a conventional beam and try to compensate for the changes by adjusting the machine output such as angle steering magnets, buncher steering magnets, position steering magnets, electron gun output, the pulse structure within the accelerator waveguide, and radiofrequency power. These automatic adjustments could be detrimental to the delivery of FLASH beams, so the Steering servo and Dose rate servo mechanisms were switched to off position in the electronics cabinet of the treatment console area (where the energy programing cards were located). Also, in the treatment console interface (supplementary Fig A1), via "Trigger Enable", the Dose servo was off and the Motor Pot Gun was turned on. AFC (automatic Frequency Control) and PFN (Pulse Forming Network) servos were kept on. The lack of steering, dose, and dose rate servos during FLASH delivery caused many interlocks in the treatment console, which were discussed further in the next section.

**6.     Interlocks**

The modification to the Linear accelerator and the servo caused interlock alerts associated with the carousel, air valves, target position, energy switch, and dose, dose rate, and beam steering servos. A figure with the resultant treatment console interface was included in supplementary figure A1. All the red highlighted interlocks were associated with the manual modifications of the LINAC structure and were overridden. The black print, blue highlighted interlocks were monitored by the two ionization chambers in the gantry head and were overridden since the beam was no longer in normal conditions. Without active monitoring of the beam with the ion chamber, alternative methods of characterizing the beam were required and discussed in section C (FLASH Dosimetry).

**7.    Conversion Back to Clinical Configuration**

The conversion back to clinical conventional dose rate beam was achieved by executing the steps from Table 1 in reverse. Dose, dose rate, and beam steering servos were turned back on in the electronics cabinet. The dose servo and motor pot gun, via "Trigger Enable", was verified to be on and off, respectively (the treatment console automatically did this after logging off service mode but required verification). The constraints on the target and energy switch motion while in FLASH positions were removed, and the lead blocks were restored in the gantry head. The empty port was re-plugged to return the LINAC to its designed safe operating mode. The air pressure connected to the air valves was turned back on and the cover for the LINAC gantry was mounted back. The user could log off service mode, log back in, select a desired energy beam, to ensure all interlocks were no longer there. Beam characteristics (dose, axial/positional symmetry, flatness, energy, position, and beam size) for all clinically commissioned beam energies were verified to be in normal conditions using a Sun Nuclear (Melbourne, FL) Daily QA phantom.

**B.    Pulse Control**

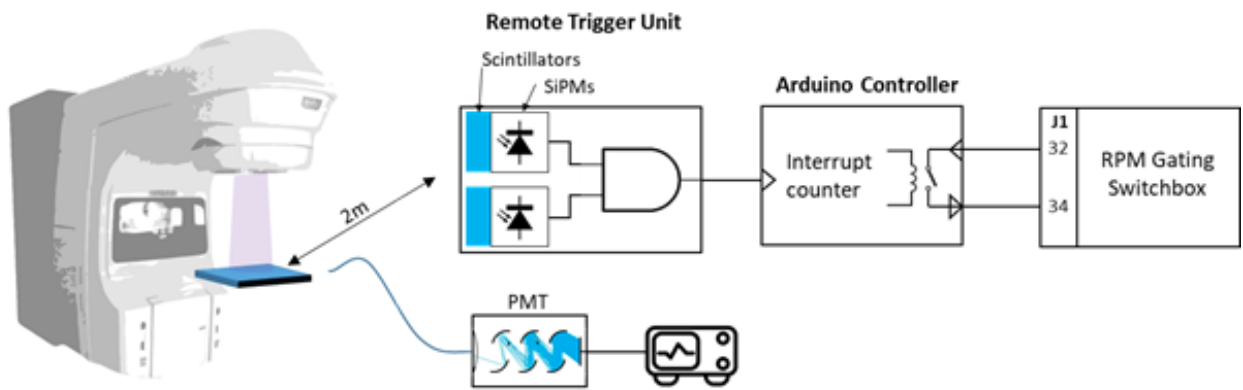

**Figure 2**. Schematic overview of the pulse control and monitoring circuitry.

Precise FLASH pulse delivery requires a control circuit that will gate on and off the electron beam and includes an independent pulse-counting feedback detector. Adapting and improving the approach of Schüler et al ,[13] we developed an Arduino Mega 2560 (Arduino LLC, MA) control circuit which turned on a gating reed relay for a period of time necessary to deliver predefined number of pulses, or equal to pre-defined timeout, whichever runs out first. The pulse-counting routine was programmed as interrupt-based, to avoid pulse skipping due to relatively long execution time of the ancillary routines in Arduino firmware (display, keyboard). Pulses were counted remotely using a coincidence-based scattered radiation detector (DoseOptics LLC, NH),[23] which included two scintillators and silicon photomultipliers. Coincident detection minimized the chance of counting a spurious high energy photon/particle or cosmic ray as an actual FLASH pulse. The output from reed relay was connected to pins 32 and 34 of the gating switchbox (Varian Inc., CA), effectively asserting an MLC hold-off signal and thereby enabling pulses only when the relay is on.

### C. FLASH Dosimetry

#### 1. Passive Dosimetry Using Gafchromic Film

Dosimetry at high dose per pulse conditions is not straightforward.[24] Most dosimeters tend to show dose-rate dependence and saturation effects at such high dose-rates. Due to the excellent dose-rate and energy independence[19] ranging from CONV (0.078 Gy/s) to ultra-high dose rates (>15 x $10^9$ Gy/s),[25] Gafchromic Film EBT-XD (Lot: 04282001) (Ashland Inc., Covington, KY) was chosen as the reference dosimeter in this study. A calibration curve relating change in optical density to dose was obtained at conventional dose rates for a 9 MeV beam, by delivering known doses ranging from 10 cGy upto 6000 cGy to 8 different slices of film. Since radiochromic film has been reported to be energy independent to within 0.5% for the energy range 6 MeV to 18 MeV, the calibration curve obtained for a 9 MeV beam should be valid for the FLASH beam. The film was digitized using an Epson 11000XL flatbed scanner (Suwa, Japan) at a resolution of 96 dots per inch (DPI) with 48-bit depth (16 bit per color channel). The machine's ability to deliver FLASH dose rate and other important beam parameters such as percentage depth dose, beam profiles and beam penumbra were quantified using data obtained from film. All dosimetric measurements were performed at iso-center, unless otherwise stated. For depth dose profiles, a film was placed parallel to the beam's axis (i.e vertically) between two 5 cm solid water slabs at isocenter. For assessing field homogeneity and size, a large piece of film was placed at the SSD=100 cm with no build up (depth = 0cm) and 5 cm of solid water underneath to provide backscatter. To quantify mean dose-rate and dose per pulse, a fixed number of pulses were delivered to the film placed at iso-center and the total dose was recorded. The total dose, number of pulses and the repetition rate of the LINAC can then be used to calculate important FLASH parameters, such as average dose-rate, and dose per pulse. Additionally, to verify dose measurements with film, another dose-rate independent dosimeter, OSLD,[25] was calibrated and irradiated with the FLASH beam.

## 2. Real Time Dosimetry Using Cherenkov Emission

To assess beam stability and variation in dose delivered per pulse in real-time, a HC-120 series photomultiplier tube (PMT) (Hamamatsu, Shizuoka, Japan) was coupled to a fiber optic. The fiber optic produces Cherenkov emission in response to radiation. The optical fiber was placed adjacent to the radiation field for all experiments. This allowed Cherenkov emission sensing on a pulse by pulse basis and assess the stability of beam output with respect to the number of pulses delivered. Since Cherenkov emission was instantaneous and the PMT had a dead-time on the order of a few nanoseconds, real-time output verification using this technique was feasible. The output from the PMT was fed into a BNC cable connected to a 3000 series PicoScope (PicoTechnology, UK). To be able to fully resolve single pulses in real-time, the PicoScope was interfaced with MATLAB via the instrument toolbox. This allowed leverage over the segmented memory feature of the PicoScope and store a large number of individual pulses in the memory buffer, at a high sampling resolution (2 ns), without running into memory storage problems. The data acquired with PMT in real-time was compared against digitized film readings post irradiation.

## III. Results and Discussion

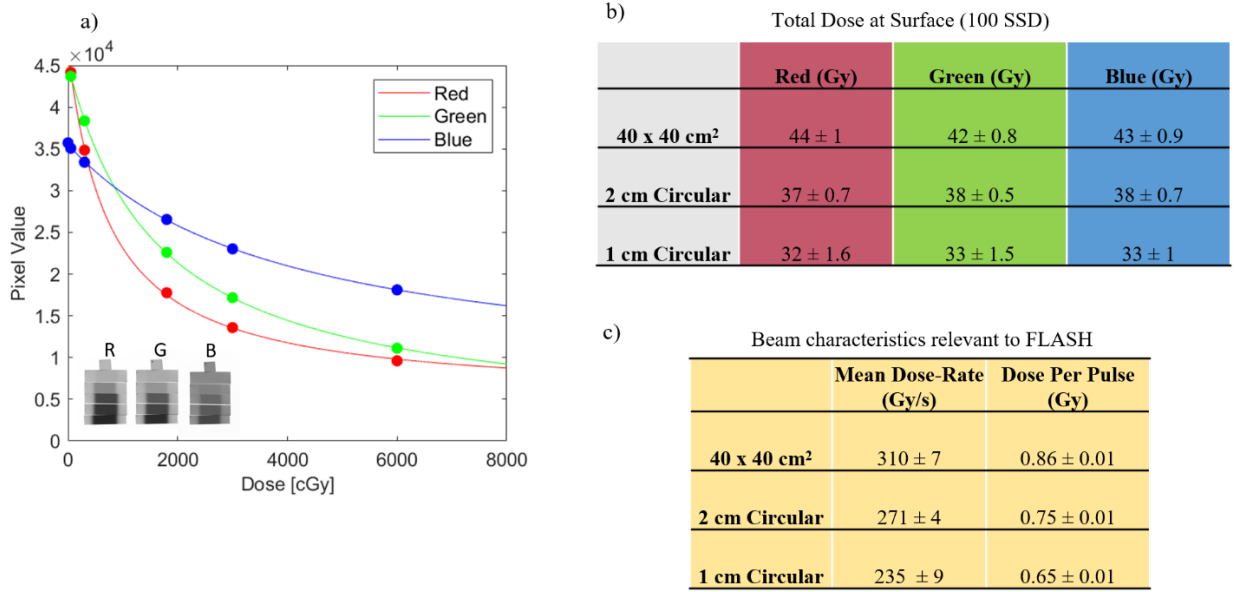

**Figure 3. a.** Calibration curve for Gafchromic Film EBT-XD. The inset shows the calibration films for the RGB channels **b.** Total dose delivered over 50 pulses to film positioned at 100 SSD, for different beam sizes. The dose recorded in the three different channels agree well with each other **c.** Beam characteristics pertinent to FLASH-RT. Mean dose-rate is calculated for a pulse repetition rate of 360 Hz.

### A. Dose-Rate Verification

The calibration curve relating the digitized film pixel values to dose is shown in Figure 3a. The calibration procedure and the fitting function proposed by Lewis et al was used.[26] Additionally, the EBT-XD film was preferred over the widely used EBT-3 Gafchromic film, due to the increased dynamic range the former offers over EBT-3. EBT-XD also mitigates other uncertainties associated with film dosimetry such as orientation effects and lateral response artefacts.[27,28] The dose recorded in the red, green and blue channels for a total of 50 pulses is shown in Figure 3b. Additional information about the fitting procedure for film is described in the supplementary document. The dose reported by the three channels was within 1%, so the average dose value of the three channels was used for all subsequent experiments. Beam characteristics, relevant to FLASH-RT are shown in Figure 3c. The dose per pulse was calculated by dividing the total dose recorded by the film by the number of pulses delivered. For mean dose-rate, a pulse repetition rate of 360 Hz (600 MU/min) was used. With the jaws completely open (field size of 40 cm² at iso-center), a dose per pulse of **0.86 ± 0.01 Gy** was achieved. This translates to a mean dose-rate (**310 Gy/s**), which is above the reported threshold of 40 Gy/s needed to elicit the FLASH effect. For pre-clinical, animal studies, smaller field sizes are of more interest compared to wide field sizes. To quantify the Clinic's ability to deliver FLASH at small field sizes, a 6 x 6 cm$^2$ electron applicator was used with circular 1 cm and 2 cm cut-outs. The mean dose-rate was again above the 40 Gy/s threshold for all field sizes (shown in Fig 3b).

To verify our measurements with film, an OSLD was also placed along with film on the central axis of a broad 40 x 40 cm2 FLASH beam. The two dosimeters in this setup were placed at a depth of 4 cm with 100 cm SSD and a total of 30 pulses were delivered. The dose recorded by the OSLD and film (average of the three color channels) was $18.5 \pm 0.06$ Gy and $18.2 \pm 0.03$ Gy, respectively.

**B.     Beam Characteristics**

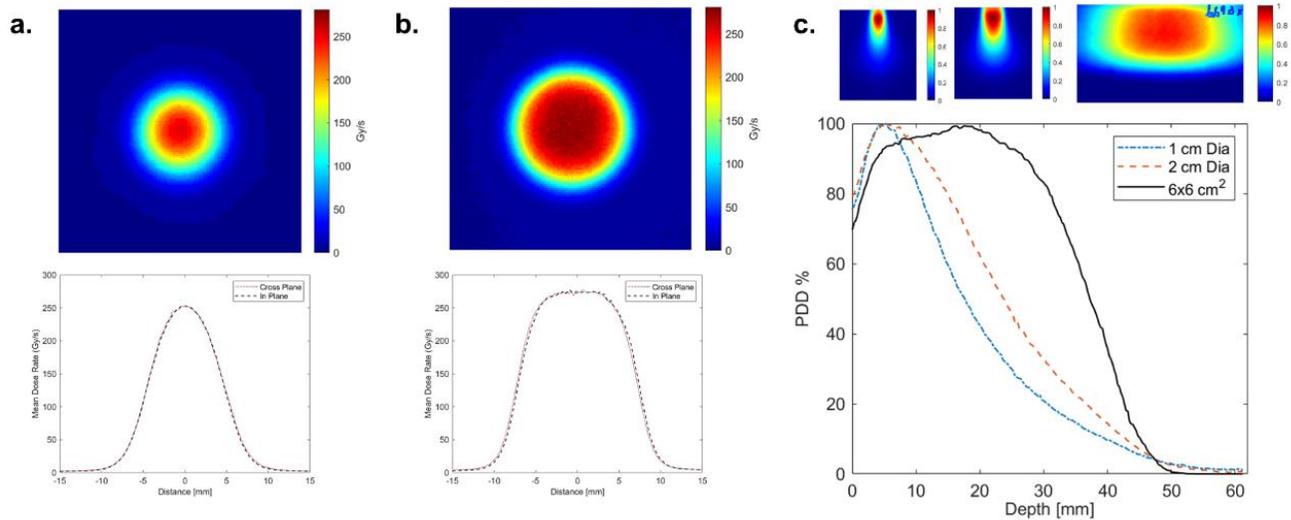

**Figure 4. a and b.** Surface profile recorded by film placed at 100 SSD for the 1 cm and 2 cm circular beam, respectively. The second row shows the central axis cross and in-line profiles for both beams. **c.** Central axis depth dose curves for the 1 cm circular, 2 cm circular and 6 cm$^2$ square beam. Above the graph, the digitized and calibrated film images are displayed for the three beams. (Left to right: 1 cm, 2 cm, 6 cm$^2$).

Important beam characteristics such as field size, homogeneity, depth dose curves were measured with film. Particularly, surface profiles were obtained with film perpendicular to the beam axis at 100 SSD. Depth dose curves were obtained with film aligned vertically (i.e parallel to the beam axis) between two 5 cm SW slabs, again at 100 SSD. Three beams were used for this experiment; 1) 1 cm circular, 2) 2 cm circular and

3) 6 x 6 cm$^2$ beam collimated via an electron applicator. The 2D surface dose distribution, along with central axis beam profiles for the 1 cm and 2 cm beam are shown in Figure 4a and 4b. The cross and in-line profiles matched well with each other, indicating excellent radial symmetry. The mean dose-rate on the central axis was recorded to be 235 ± 9 Gy/s and 271 ± 4 Gy/s for the 1 cm and 2 cm beam, respectively. Central axis percentage depth dose curves (PDD) for the three beams are shown in Figure 4c. The practical range ($R_p$) obtained from the measured PDD agrees with that of a typical 10 MeV electron beam (~ 50 mm). Additionally, as expected, the point of maximum dose ($d_{max}$) moved closer to the surface and the surface dose increased with decreasing field size. The depth dose curve for electrons is highly dependent on the angle of incidence of the incident beam, and therefore requires careful alignment of the film to the beam axis. In addition, air gaps between the two slabs can also cause errors. Therefore, surface dose measurements should be used with caution, but beyond the surface, the current setup can be used to obtain depth dose curves that can accurately characterize beam quality metrics such as $R_p$, $R_{50}$, $R_{80}$, and $R_{90}$ etc.

## C. Temporal Stability

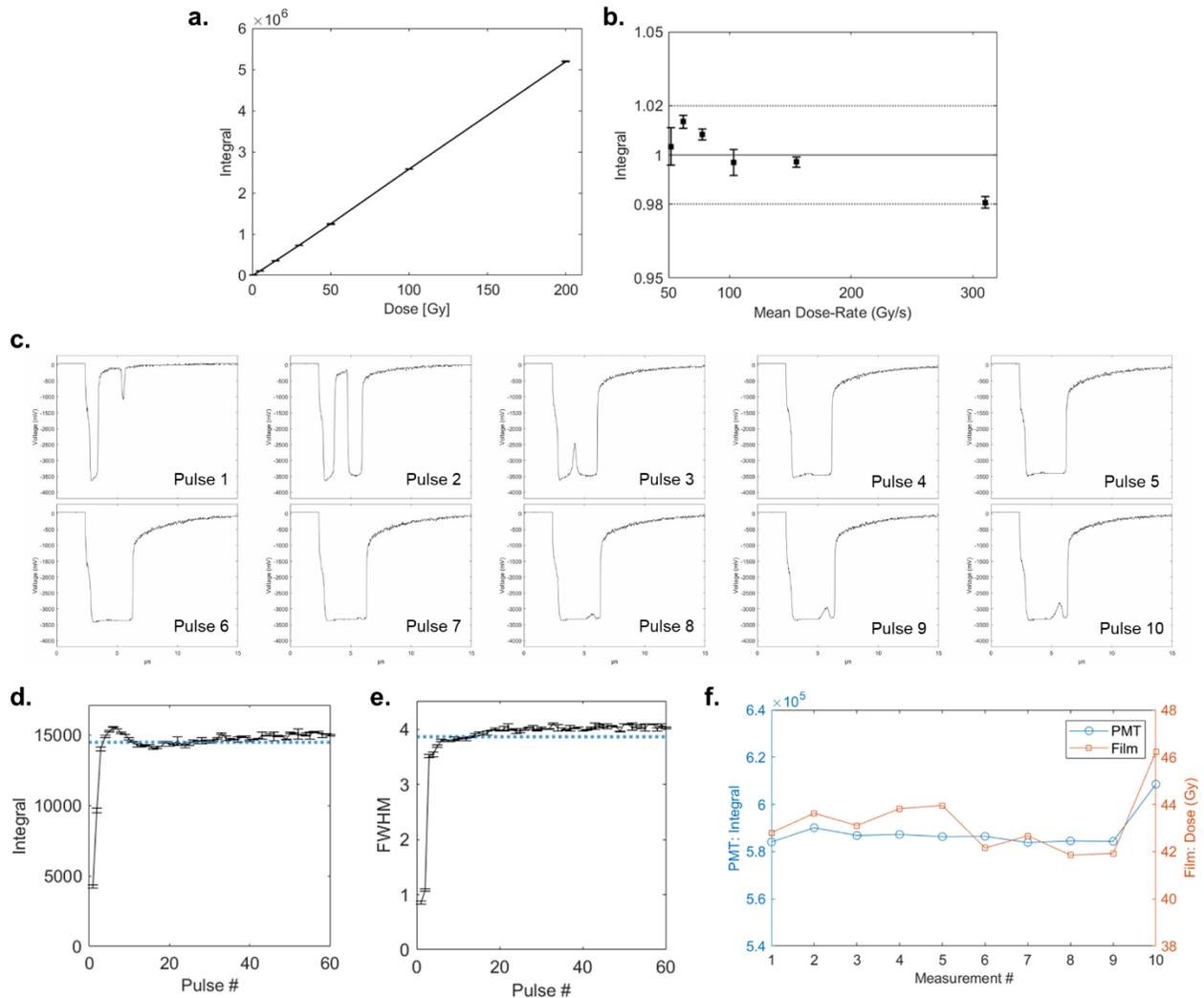

**Figure 5 a.** Linearity of the area under the curve of the Cherenkov signal detected with PMT with increasing dose ($R^2 = 0.99$). The error bars are contained within the linear line plotted on the graph and represent standard deviation from 3 repeated measurements **b.** The PMT response to the same number of pulses (50), but at varying dose-rates. **c.** Montage of the temporal profile for the first 10 pulses. **d and e.** Pulse by pulse stability of the beam by showing the variation of area under the curve (integral) and full-width at half maximum (FWHM), respectively, for a sequence of 60 pulses. The error bars in d and e represent standard error over 3 repeated measurements. **f.** Long term stability of the FLASH beam assessed with the PMT and film measurements by delivering 50 pulses repeatedly at different time intervals (3 minutes between each acquisition).

While film was useful for absolute dosimetry, the lack of real-time feedback was a problem that still needed to be solved for accurate and reproducible dose delivery at FLASH dose rates. Thus, a PMT coupled to fiber optic was used to assess the temporal stability of the beam on a pulse by pulse basis. To assess the ability of the PMT for real-time dose read-out, dose and dose-rate linearity had to be established. The former was established by delivering an increasing number of pulses and the latter was established by delivering the same number of pulses (i.e. same dose) at different repetition rates. This data was presented in Figure 5a and 5b. The integral of the Cherenkov pulse intensity varied linearly with dose at a wide range of doses ($R^2$ = 0.99). Additionally, as shown in Figure 5b, the signal obtained from the PMT was independent of any dose-rate effects. The signal reported by the PMT was within 2% for the same dose delivered at different dose-rates (50-300 Gy/s). Once dose and dose-rate linearity were established, the PMT was used to monitor beam output in real-time on a pulse by pulse basis. The full-width at half maximum (FWHM) and the area under each pulse was monitored for a 60 pulse delivery. The results are shown in Figure 5d and 5e. The initial 4-6 pulses represented the beam 'ramp-up' time and the output stabilized after >10 pulses, which can be seen in the temporal profiles of the first 10 pulses in figure 5c. In particular, the first pulse delivers an output which was on average 30 % lower than the average dose per pulse value. The FWHM metric followed a similar trend and stabilized to the expected pulse width of 4 μs after ~10 pulses. The same trend was observed for all repetition rates available on the LINAC (i.e 60 Hz to 360 Hz). The temporal width assessment using PMT can prove to be a useful tool if beam tuning is required to optimize dose-rate. Traditionally, beam tuning is performed based on the measured pulse width of the target current (TargI) pulse. For a machine operating in resonance (i.e. electrons are injected in the waveguide at a time such that they always experience an accelerating potential through RF wave), the typical pulse width of the TargI signal is about 4-5 μs. Any sort of mistuning will show up as different variations in the TargI pulse. Since, in this study, the Target was removed, the TargI signal was no longer meaningful. In this case, if the beam output were to be maximized the signal from the PMT could serve as a surrogate for the TargI signal. In addition to the pulse by pulse stability, the long-term stability of the machine output was also tested by delivering the same number of pulses (50) at 10 time points, three minutes apart. The results are shown in Figure 5f. The standard deviation for the 10 measurements for film and the PMT was, 3.1% and 1.3% respectively. Interestingly, the last data point (at ~ 30 minutes) represented a significant increase in dose when compared to the previous 9 measurements. Looking at the pulse by pulse variation for this acquisition, it was seen that the ramp-up period had significantly been reduced i.e. the first pulse delivered a dose which was 90% of the average dose per pulse. The FWHM of the first pulse in this case was also found to be ~3.5 μs as opposed to the 1 μs pulse that was observed for the first 9 measurements. Importantly, this increase in dose was reflected by both film and the PMT.

**D.    Conversion to Conventional Beam**

The LINAC was converted back to deliver conventional clinical beams in under 20 minutes by turning on the air drive and removing the restrictions on the carousel, energy switch, and target that prevented automatic motions. The beam characteristics for all clinical conventional beam energies measured by the Daily QA phantom was shown in supplementary figure A2. The dose difference measured in comparison to baseline (measurements prior to LINAC modifications) were within ~1%. The axial and position symmetry were within the normal 2% tolerance. The electron or photon energy measurements were within 0.1%. The inline and crossline shift and size of the beams were generally within 1mm. All the measurements were within the normal tolerance of baseline, deeming the machine safe for clinical use.[29]

**IV.    Conclusions**

In this study, a clinical linear accelerator was converted from delivering conventional dose rate (~0.1 Gy/s) photon beams to FLASH dose rate (>300 Gy/s) electron beams. Such high dose rates were achieved at the isocenter (100 cm source to surface distance) by retracting x-ray target and traversing the electron beam through an empty carousel port with no FF, SF, or port cover. Guidelines for conversion were explicitly described and details on the modification effect of the linear accelerator components were provided. The methods of conversion readily allowed for efficient conversion back to clinical conventional electron and photon beams. The LINAC was verified based on quality assurance protocols and measuring beam output, energy, flatness, symmetry, and field size. FLASH dose rates of >300 Gy/s were achieved at isocenter based on Gafchromic Film EBT-XD measurements. With reduced SSD and further tuning of the beam output, dose rates upwards of 600 Gy/s can be achieved. Film based measurements were compared against OSLD, and the two dosimeters agreed to within 1 %. In addition to passive film dosimetry, real-time dosimetry beam monitoring was performed with a Cherenkov based PMT detector. Since, the transmission chamber in the gantry head can suffer from ion-recombination and saturation effect at such high doses per pulse, the Cherenkov based detector can play the role of the traditional ionization transmission chamber by providing feedback to the LINAC in real-time for FLASH beam delivery. A ramp-up period, which lasted the first 4-5 pulses, was observed with the Cherenkov detector. For preclinical animal studies, this implies that for reproducible dose per pulse delivery upwards of 10 pulses might be required. To properly study the mechanisms underlying FLASH-RT, it is important that single pulses be delivered reproducibly and accurately. It was observed that the first few pulses, even though considerably lower than the average dose per pulse, were reproducible and still in the FLASH regime (~0.1-0.3 Gy/Pulse). Additionally, the long term stability (> 30 mins) of the machine output was confirmed with PMT and film data. Therefore, the machine can be faithfully used to deliver doses for animal studies over extended periods of time. Other clinically used dosimeters were also employed in this study. A commercial ionization chamber was found to saturate, even for the low dose first pulse. Dosimetry under FLASH conditions remains a challenging task; dosimetric tools that can offer real-time, pulse resolved, dose-rate independent dosimetry are required so that machine output can be monitored and controlled in real-time.

## V.     Acknowledgments

This work was supported by the Norris Cotton Cancer Center seed funding through core grant P30 CA023108 and through seed funding from the Thayer School of Engineering, as well as support from grant R01 EB024498. The authors are grateful for discussions with Acceleronics Inc. about aspects of this work. We acknowledge informative and helpful conversations with Emil Schüler, PhD., Kristoffer Petersson, PhD., and Tania Karan, MSc.

# References

1. Hornsey S, Alper T. Unexpected Dose-rate Effect in the Killing of Mice by Radiation. *Nature*. 1966;210(5032):212-213. doi:10.1038/210212a0
2. Hornsey S, Bewley DK. Hypoxia in Mouse Intestine Induced by Electron Irradiation at High Dose-rates. *Int J Radiat Biol Relat Stud Phys Chem Med*. 1971;19(5):479-483. doi:10.1080/09553007114550611
3. Favaudon V, Fouillade C, Vozenin M-C. Radiothérapie « flash » à très haut débit de dose : un moyen d'augmenter l'indice thérapeutique par minimisation des dommages aux tissus sains ? *Cancer/Radiothérapie*. 2015;19(6-7):526-531. doi:10.1016/j.canrad.2015.04.006
4. Montay-Gruel P, Acharya MM, Petersson K, et al. Long-term neurocognitive benefits of FLASH radiotherapy driven by reduced reactive oxygen species. *Proc Natl Acad Sci*. 2019;116(22):10943-10951. doi:10.1073/pnas.1901777116
5. Favaudon V, Caplier L, Monceau V, et al. Ultrahigh dose-rate FLASH irradiation increases the differential response between normal and tumor tissue in mice. *Sci Transl Med*. 2014;6(245):245ra93-245ra93. doi:10.1126/scitranslmed.3008973
6. Vozenin M-C, De Fornel P, Petersson K, et al. The Advantage of FLASH Radiotherapy Confirmed in Mini-pig and Cat-cancer Patients. *Clin Cancer Res*. 2019;25(1):35-42. doi:10.1158/1078-0432.CCR-17-3375
7. Bourhis J, Sozzi WJ, Jorge PG, et al. Treatment of a first patient with FLASH-radiotherapy. *Radiother Oncol*. 2019;139:18-22. doi:10.1016/j.radonc.2019.06.019
8. Bazalova-Carter M, Esplen N. On the capabilities of conventional x-ray tubes to deliver ultra-high (FLASH) dose rates. *Med Phys*. 2019;46(12):5690-5695. doi:10.1002/mp.13858
9. Patriarca A, Fouillade C, Auger M, et al. Experimental Set-up for FLASH Proton Irradiation of Small Animals Using a Clinical System. *Int J Radiat Oncol*. 2018;102(3):619-626. doi:10.1016/j.ijrobp.2018.06.403
10. Buonanno M, Grilj V, Brenner DJ. Biological effects in normal cells exposed to FLASH dose rate protons. *Radiother Oncol*. 2019;139:51-55. doi:10.1016/j.radonc.2019.02.009
11. Darafsheh A, Hao Y, Zwart T, et al. Feasibility of proton FLASH irradiation using a synchrocyclotron for preclinical studies. *Med Phys*. Published online June 15, 2020. doi:10.1002/mp.14253
12. Eling L, Bouchet A, Nemoz C, et al. Ultra high dose rate Synchrotron Microbeam Radiation Therapy. Preclinical evidence in view of a clinical transfer. *Radiother Oncol*. 2019;139:56-61. doi:10.1016/j.radonc.2019.06.030
13. Schüler E, Trovati S, King G, et al. Experimental Platform for Ultra-high Dose Rate FLASH Irradiation of Small Animals Using a Clinical Linear Accelerator. International Journal of Radiation Oncology*Biology*Physics. 2017;97(1):195-203. doi:10.1016/j.ijrobp.2016.09.018
14. Loo BW, Schuler E, Lartey FM, et al. (P003) Delivery of Ultra-Rapid Flash Radiation Therapy and Demonstration of Normal Tissue Sparing After Abdominal Irradiation of Mice. *Int J Radiat Oncol*. 2017;98(2):E16. doi:10.1016/j.ijrobp.2017.02.101
15. Lempart M, Blad B, Adrian G, et al. Modifying a clinical linear accelerator for delivery of ultra-high dose rate irradiation. *Radiother Oncol*. 2019;139:40-45. doi:10.1016/j.radonc.2019.01.031
16. Karzmark CJ. *A Primer on Theory and Operation of Linear Accelerators in Radiation Therapy*. 3rd edition. Medical Physics Pub; 2017.
17. Anderson R, Lamey M, MacPherson M, Carlone M. Simulation of a medical linear accelerator for teaching purposes. *J Appl Clin Med Phys*. 2015;16(3):359-377. doi:10.1120/jacmp.v16i3.5139
18. Khan FM, Gibbons JP. *Khan's the Physics of Radiation Therapy*. Fifth edition. Lippincott Williams & Wilkins/Wolters Kluwer; 2014.

19. Sharma S. Unflattened photon beams from the standard flattening filter free accelerators for radiotherapy: Advantages, limitations and challenges. *J Med Phys*. 2011;36(3):123. doi:10.4103/0971-6203.83464
20. Zhang R, Rahman M, Ashraf MR, et al. Commissioning of the First Treatment Planning System for Electron Flash Radiation Therapy in a Clinical Setting. In: ; Under Review.
21. Van Dyk J, ed. *The Modern Technology of Radiation Oncology: A Compendium for Medical Physicists and Radiation Oncologists*. Medical Physics Pub; 1999.
22. Attix FH. *Introduction to Radiological Physics and Radiation Dosimetry*. Wiley; 1986.
23. Ashraf MR, Bruza P, Krishnaswamy V, Gladstone DJ, Pogue BW. Technical Note: Time-gating to medical linear accelerator pulses: Stray radiation detector. *Medical Physics*. Published online December 14, 2018. doi:10.1002/mp.13311
24. Ashraf MR, Rahman M, Zhang R, Williams BB, Gladstone DJ, Bruza P. Dosimetry for FLASH Radiotherapy: A Review of Tools and the Role of Radioluminescence and Cherenkov Emission. :26.
25. Karsch L, Beyreuther E, Burris-Mog T, et al. Dose rate dependence for different dosimeters and detectors: TLD, OSL, EBT films, and diamond detectors: Dose rate dependence for different dosimeters and detectors. *Medical Physics*. 2012;39(5):2447-2455. doi:10.1118/1.3700400
26. Lewis D, Micke A, Yu X, Chan MF. An efficient protocol for radiochromic film dosimetry combining calibration and measurement in a single scan. *Med Phys*. 2012;39(10):6339-6350. doi:10.1118/1.4754797
27. Khachonkham S, Dreindl R, Heilemann G, et al. Characteristic of EBT-XD and EBT3 radiochromic film dosimetry for photon and proton beams. *Phys Med Biol*. 2018;63(6):065007. doi:10.1088/1361-6560/aab1ee
28. Lewis DF, Chan MF. Technical Note: On GAFChromic EBT-XD film and the lateral response artifact: Lateral response corrections for GAFChromic EBT-XD film. *Med Phys*. 2016;43(2):643-649. doi:10.1118/1.4939226
29. Klein EE, Hanley J, Bayouth J, et al. Task Group 142 report: Quality assurance of medical accelerators a): Task Group 142 Report: QA of Medical Accelerators. *Med Phys*. 2009;36(9Part1):4197-4212. doi:10.1118/1.3190392

**Supplementary Material A**

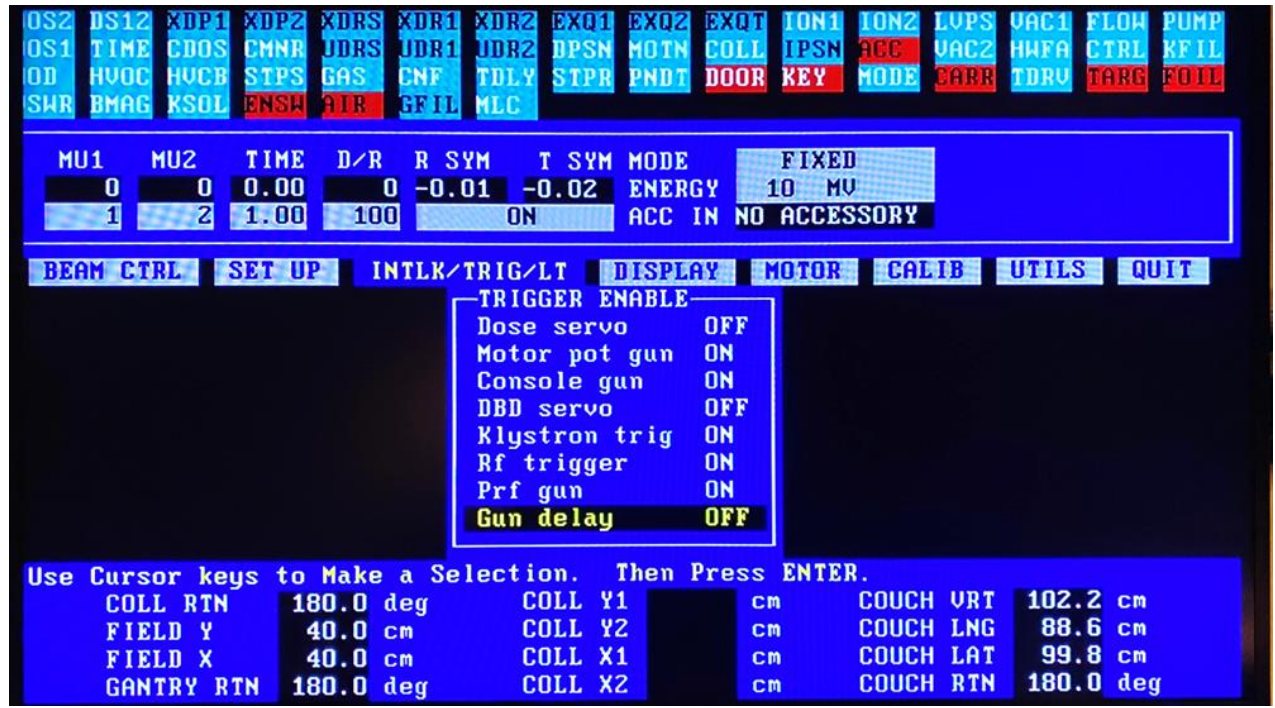

**Figure A1.** Treatment Console interface when LINAC was converted to deliver FLASH beams. Interlocks were overridden (black font) and interlocks warnings (highlighted red) remained due to the manual modifications. "Trigger Enable" settings are included to turn off Dose and Dose rate servo.

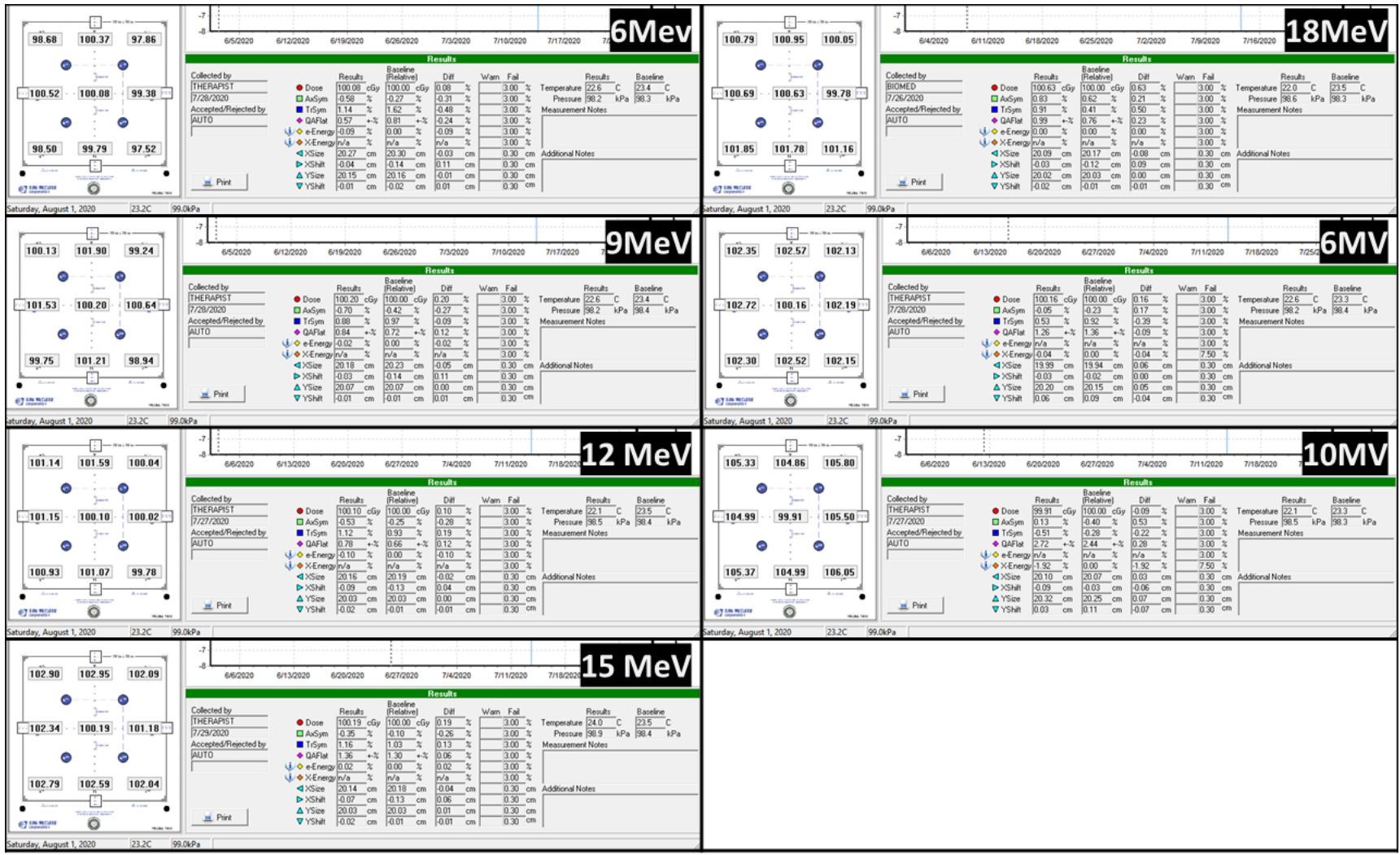

**Figure A2.** Dose, axial/positional symmetry, flatness, energy, position, and beam size results for all commissioned energies after conversion back to conventional beam.

# Appendix: Film Dosimetry

$$\text{Pixel Value} = a + \frac{b}{D - c}$$

The change in optical density of the film was related to absorbed dose through the fitting function described above. a, b and c are fitting constants. D is the absorbed dose. The rational function is recommended because it corresponds well to the physical dose-response characteristics of the radiochromic film (i.e exponential darkening of the film in response to radiation). Traditionally used polynomial functions can result in oscillatory behavior at high doses and should therefore be avoided when conducting film dosimetry at high doses relevant in FLASH (*). All image processing was performed on the MATLAB (MathWorks, Natick, MA) computing environment. The rational function was fitted to the red, green and the blue channel. The calibration curves were applied to their respective color channels and dose reported by the three channels was compared. If the dose reported by the three channels agreed within 3%, an average of the three channels was used as the final dose reading. The fitting parameters for the EBT-XD film are given below:

|   | R | G | B |
| --- | --- | --- | --- |
| a | 5167 | 1890 | 6744 |
| b | 3.14e+05 | 7.1e+05 | 1.13e +06 |
| c | -7.5 | -16.4 | -39.2 |

The goodness of the fit for each color channel is given below:

|   | SSE | R-square | RMSE |
|---|---|---|---|
| **R** | 2.5e+05 | 0.9998 | 285.8 |
| **G** | 1.0e+05 | 0.9999 | 184.2 |
| **B** | 5.6e+04 | 0.9998 | 138.5 |

\*.    Lewis D, Micke A, Yu X, Chan MF. An efficient protocol for radiochromic film dosimetry combining calibration and measurement in a single scan: Efficient protocol for radiochromic film dosimetry. Med Phys. 2012 Sep 27;39(10):6339–50.